\documentclass[useAMS,usenatbib]{mn2e}
\usepackage{amsmath,amssymb,wasysym,graphicx,epsfig}

\title[]{The discovery of a very cool, very nearby brown dwarf in the Galactic plane}

\author[Lucas et al.]{P. W. Lucas$^1$,
C.G. Tinney$^2$, Ben Burningham$^1$, S. K. Leggett$^3$, David J. Pinfield$^1$,
\newauthor Richard Smart$^4$, Hugh R.A. Jones$^1$, Federico Marocco$^4$, Robert J. Barber$^5$,
\newauthor Sergei N. Yurchenko$^6$, Jonathan Tennyson$^5$, Miki Ishii$^7$, Motohide Tamura$^8$,
\newauthor Avril C. Day-Jones$^9$, Andrew Adamson$^{10,3}$, France Allard$^{11}$, Derek Homeier$^{12}$\\
$^1$Centre for Astrophysics Research, University of Hertfordshire, College Lane, 
Hatfield AL10 9AB, UK\\
$^2$Dept of Astrophysics, University of New South Wales, 2052, Australia\\
$^3$Gemini Observatory, Northern Operations Centre, 670 North AÕohoku Place, Hilo HI 96720, USA\\
$^4$INAF/Osservatorio Astronomico di Torino, Strada Osservatorio 20, 10025 Pino Torinese, Italy\\
$^5$Dept of Physics and Astronomy, University College London, London WC1E 6BT, UK\\
$^6$Technische Universit\"{a}t Dresden, Institut f\"{u}r Physikalische Chemie und Elektrochemie, D-01062 
Dresden, Germany\\
$^7$Subaru Telescope, National Astronomical Observatory of Japan, 650 N.A'ohoku Place, Hilo, HI 96720, USA\\
$^8$National Astronomical Observatory of Japan, 2-21-1 Osawa, Mitaka, Tokyo 181-8588, Japan\\
$^9$Universidad de Chile, Camino el Observatorio, \#1515, Las Condes, Santiago, Chile, Casilla 36-D, Chile\\
$^{10}$Joint Astronomy Centre, 660 N.A'ohoku Place, University Park, Hilo, HI 96720, USA\\
$^{11}$CRAL, Universite de Lyon, \'{E}cole Normale Sup\'{e}rieur, 46 allee d'Italie, 69364 Lyon Cedex 07, 
France\\
$^{12}$Institut f\"{u}r Astrophysik G\"{o}ttingen, Georg-August-Universit\"{a}t, Friedrich-Hund-Platz 1, 
D-37077 G\"{o}ttingen, Germany}

\begin{document}
\newfont{\ssq}{cmtt10} 

\date{Accepted 2010. Received 2010; in original form June 2010}

\pagerange{\pageref{firstpage}--\pageref{lastpage}} \pubyear{2010}

\maketitle

\label{firstpage}

\begin{abstract}
We report the discovery of a very cool, isolated brown dwarf, UGPS 0722-05, with the 
UKIDSS Galactic Plane Survey. The near-infrared spectrum displays deeper H$_2$O
and CH$_4$ troughs than the coolest known T dwarfs and an unidentified 
absorption feature at 1.275~$\mu$m. We provisionally classify the object as a T10 dwarf but note 
that it may in future come to be regarded as the first example of a new spectral type. The distance 
is measured by trigonometric parallax as $d$=4.1$_{-0.5}^{+0.6}$~pc, making it the closest known
isolated brown dwarf. With the aid of {\it Spitzer}/IRAC we measure $H$-[4.5] = 4.71. 
It is the coolest brown dwarf presently known -- the only known T dwarf that is redder
in $H$-[4.5] is the peculiar T7.5 dwarf SDSS J1416+13B, which is thought to be warmer and more 
luminous than UGPS 0722-05.
Our measurement of the luminosity, aided by Gemini/T-ReCS $N$ band photometry, is 
L=9.2$\pm 3.1 \times10^{-7}$L$_{\odot}$. Using a comparison with well studied T8.5 and T9 dwarfs 
we deduce T$_{eff}$=520$\pm$40~K.
This is supported by predictions of the Saumon \& Marley models. With 
apparent magnitude $J=16.52$, UGPS 0722-05 is the brightest brown dwarf discovered by UKIDSS so 
far. It offers opportunities for future study via high resolution near-infrared spectroscopy 
and spectroscopy in the thermal infrared.
\end{abstract}


\section{Introduction}
In the past 15 years $\sim$200 T dwarfs, have been discovered in the 
local field with the Two Micron All Sky Survey (2MASS), the Sloan Digital Sky Survey and the 
United Kingdom Infrared Telescope Deep Sky Survey (UKIDSS), see e.g. Burgasser et al. (2002); 
Leggett et al. (2000); Burningham et al. (2010a). 
These discoveries are useful for investigating the substellar mass function.
Searches in young clusters e.g. Lodieu et al. (2007); Weights et al. (2009);
Luhman (2004) suggest that the ratio of brown dwarfs to stars 
is between $\sim$1:6 and $\sim$1:3 (excluding companions) but there are 
growing indications that fewer 
exist in the local field (e.g. Burningham et al. 2010a). The nearest isolated brown dwarfs
known until now (LP944-20, DENIS-P~J025503.5-470050, DENIS~J081730.0-615520) are located 
at distances $d$=4.9-5.0~pc from the Sun (Artigau et al.2010; Tinney 1998; Costa et al. 2006).
These are the 46th, 48th and 49th nearest systems. If the mass function in the local field is
similar to that in young clusters then we should expect that some brown dwarf
primaries remain undiscovered at $d<5$~pc. 
Most such objects would have types $\ge$T8.5 in order to lie below the 2MASS detection 
limit (Skrutskie et al. 2006).

The UKIDSS Large Area Survey (LAS, Lawrence et al. 2007) has
discovered several brown dwarfs cooler than those found by 2MASS (e.g. Warren et al. 2007; 
Burningham et al. 2008 (hereafter B08); 2010b).
The CFHT Brown Dwarf Survey has yielded two similar discoveries
(e.g. Delorme et al. 2008). Most of these objects have been classified as T8.5 or T9 
dwarfs because the broad H$_2$O and CH$_4$ absorption bands in their near-infrared 
spectra are slightly deeper than those of T8 dwarfs. The slight differences in the spectra belie 
a large drop in effective temperature from 700-800~K at T8 to 500-600 K for the coolest
objects (B08; Leggett et al. 2010a, hereafter L10a). 
Here we report the discovery in the UKIDSS Galactic Plane Survey (GPS, Lucas et al. 2008, hereafter L08) 
of UGPS~J072227.51-054031.2 (hereafter UGPS~0722-05), an even cooler and less luminous brown dwarf.
The GPS is not optimal for late T dwarf searches since it employs only the $J$, $H$ 
and $K$ filters.  
For objects with negative {\it H-K} colours the GPS probes less than a quarter of
the volume of the LAS for the same area of sky. Source confusion is also a serious impediment in
a significant fraction of the GPS area.

\section{Candidate selection}

UGPS 0722-05 was identified as the only good candidate late-T or Y dwarf amongst the 604 million 
sources in GPS 6th Data Release to satisfy the colour criteria
{\it J-H}$<$-0.2 mag, {\it H-K}$<$-0.1 mag. 
We used several data quality restrictions to minimise the number of false 
candidates (see L08). The criteria were: {\ssq jmhPntErr}$<$0.3, 
{\ssq hmk\_1PntErr}$<$0.3 (limiting the uncertainty in source colours);
{\ssq jppErrbits}$<$256, {\ssq hppErrbits}$<$256, {\ssq k\_1ppErrbits}$<$256
(removes sources with photometric quality warnings); 
{\ssq pstar}$>$0.9 (requires a point-source image profile); 
{\ssq jEll}$<$0.3, {\ssq hEll}$<$0.3, {\ssq k\_1Ell}$<$0.3 (limits on ellipticity);
{\ssq $\sqrt{}$(hXi$^2$+hEta$^2$)}$<$0.3, {\ssq $\sqrt{}$(k\_1Xi$^2$+k\_1Eta$^2$)}$<$0.3
(limits on coordinate shifts between passbands).

A further constraint was to limit the search to Galactic longitudes $l>$60$^{\circ}$ and 
$l<$358$^{\circ}$. These restrictions were designed to select 
against blended stellar pairs with inaccurate photometry, which are a frequent occurrence in the most 
crowded regions of the plane. 
Only six candidates remained after this procedure. Of these, four were revealed as blended 
stellar pairs or defective data by inspection of the images and one (a candidate white dwarf) was 
ruled out by its detection in visible light in the POSS USNO-B1.0 archive. 
An image of the remaining candidate, UGPS~0722-05, is shown in Fig. 1. The coordinates measured 
on 2 March 2010 (see $\S$3) were R.A.= 07$^h$22$^m$27.29$^s$, Dec.=-05$^d$40$^m$30.0$^s$.

\section{Observations}

The Near Infrared Imager and Spectrometer (NIRI) on the Gemini North Telescope on Mauna Kea was used 
on 10th, 11th and 14th of February 2010 to take spectra covering the $J$, $K$ and $H$ 
bandpasses respectively, with total on source integration times of 16, 30 and 60 minutes respectively. 
The spectral resolution was R$\sim$500. They were reduced and calibrated with the IRAF software package 
using standard techniques, see e.g. Burningham et al. (2010a). The three spectra were then flux
calibrated using the UKIDSS photometry and combined into a single spectrum.

The UKIDSS discovery images were taken on 28 November 2006. The NIRI acquisition images obtained 3.7 
years later showed that UGPS 0722-05
has a proper motion of $\sim$1\arcsec per year. This then allowed the identification of UGPS~0722-05 
with a previously uncatalogued source in the 2MASS Atlas image acquired with the 2MASS South 
telescope on 19 October 1998. A programme of parallax observations was then begun
with the UKIRT Wide Field Camera, using the methods described in Smart et al. (2010). The object was 
observed in the $J$ passband on 19 February, 2 March, 16 March 30 March, 13 April and 27 April.
All the UKIRT images had full-width half maxima between 0.8 and 1.1\arcsec and used microstepping 
to yield a pixel scale of 0.20\arcsec. The total integration time on each occasion was 400~s.

\begin{figure}
\hspace{1.5cm} \includegraphics[scale=0.29,angle=0]{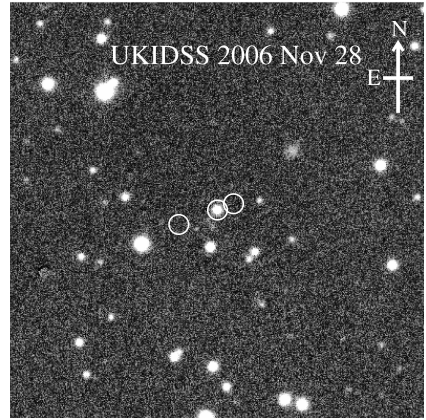}
\caption{UKIDSS $J$ band discovery image (80$\arcsec$ square). The circles illustrate the 
proper motion (see Table 3). The centre circle marks the position at the 
time of the discovery image in 2006, while the circles to the left and right mark the positions at the
time of the 2MASS image in 1998 and a UKIRT image from 2 March 2010 respectively.}
\end{figure}

\begin{table*}
\begin{minipage}{200mm} 
    \caption{Photometry of UGPS 0722-05}
    \begin{tabular}{llllccccccccccccc}
 & $i$ & $z$ & $z$ & $Y$ & $J$ & $H$ & $K$ & [3.6] & [4.5] & $L^{\prime}$ & $N$ & {\it J-H} 
& {\it H-K} \\ \hline
Magnitude & 24.80 & 20.60 & 20.42 & 17.37 & 16.52 & 16.90 & 17.07 & 14.28 & 12.19 
& 13.4 & 10.28 & -0.38 & -0.18 \\
Error & 0.13 & 0.07 & 0.06 & 0.02 & 0.02 & 0.02 & 0.08 & 0.05 & 0.04 & 0.3 & 0.24 & 0.03 & 0.08 \\
Instrument & GMOS & GMOS & WFC & WFC & WFC & WFC & WFC & IRAC & IRAC
& IRCS & T-ReCS \\
Date & 15/3 & 15/3 & 16/3 & 16/3 & 27/11 & 27/11 & 27/11 & 30/4 & 30/4 & 6/4 & 24/3;13/4 \\ 
Exposure (s) & 4800 & 120 & 400 & 120 & 80 & 80 & 40 & 1440 & 1440 & 335 & 3360 \\ 
  \hline
\end{tabular}
\end{minipage}
\end{table*}

\begin{table*}
\begin{minipage}{200mm} 
    \caption{Spectral indices for very cool brown dwarfs. Numbers in brackets are the uncertainties on the 
last 2 digits.}
    \begin{tabular}{lcccccccc}
Object & W$_J$ & H$_2$O-J & CH$_4$-J & NH$_3$-H & H$_2$O-H & CH$_4$-H & CH$_4$-K & K/J \\ \hline
UGPS 0722-05 & 0.2074(12) & 0.0339(17) & 0.1358(19) & 0.4917(21) & 0.1218(17)
    & 0.0643(13) & 0.0959(22) & 0.12615(33) \\
T9 average & 0.258 & 0.025 & 0.162 & 0.535 & 0.122 & 0.086 & 0.104 & 0.121 \\
T8 0415-09 & 0.31 & 0.030 & 0.168 & 0.625 & 0.173 & 0.105 & 0.05 & 0.134 \\
  \hline
\end{tabular}
\end{minipage}
\end{table*}

The results of multiwaveband photometry are given in Table 1. All dates refer to 2010 except for 
the UKIDSS $JHK$ photometry from 2006. The instruments listed are: GMOS (the Gemini Multi-Object 
Spectrograph on the Gemini North Telescope); WFC (the Wide Field Camera on the United
Kingdom Infrared Telescope (UKIRT)); IRAC (the Infrared Array Camera on the {\it Spitzer Space 
Telescope}; IRCS (the Infrared Camera and Spectrograph on the Subaru Telescope) and T-ReCS 
(the Thermal Region Camera Spectrograph on the Gemini South Telescope). 

All fluxes except $i$ and $z$ are Vega magnitudes. 
Data taken in the GMOS $i$ and $z$ filters and the WFC $Z$ filter were transformed to the SDSS AB 
system using the far-red spectra of 16 dwarfs ranging in type from L3 to T8. 
The large uncertainty in the IRCS $L^{\prime}$ flux is due to uncertainty in the aperture correction, 
which arose from imperfect telescope tracking. 
The T-ReCS data were taken over 2 nights. Observing conditions were somewhat variable and the signal to 
noise ratio on each night was low. The final coadded image had a signal to noise ratio of 7. The
photometric uncertainty given in Table 1 includes the uncertainties in the calibration and in the final 
aperture correction.

\section{Results}

\begin{figure}

\includegraphics[scale=0.33,angle=0]{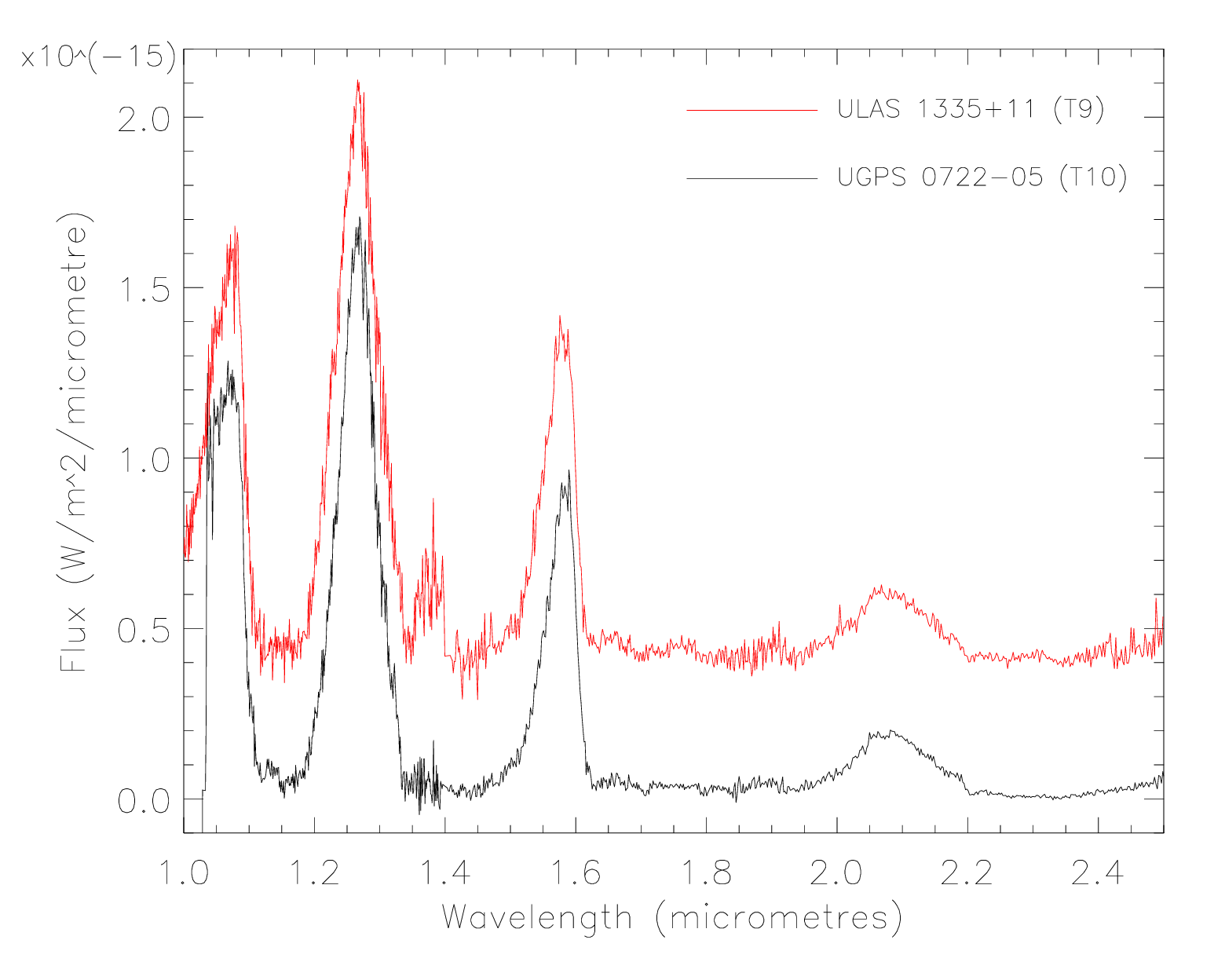}

\vspace{-4mm}
\includegraphics[scale=0.33,angle=0]{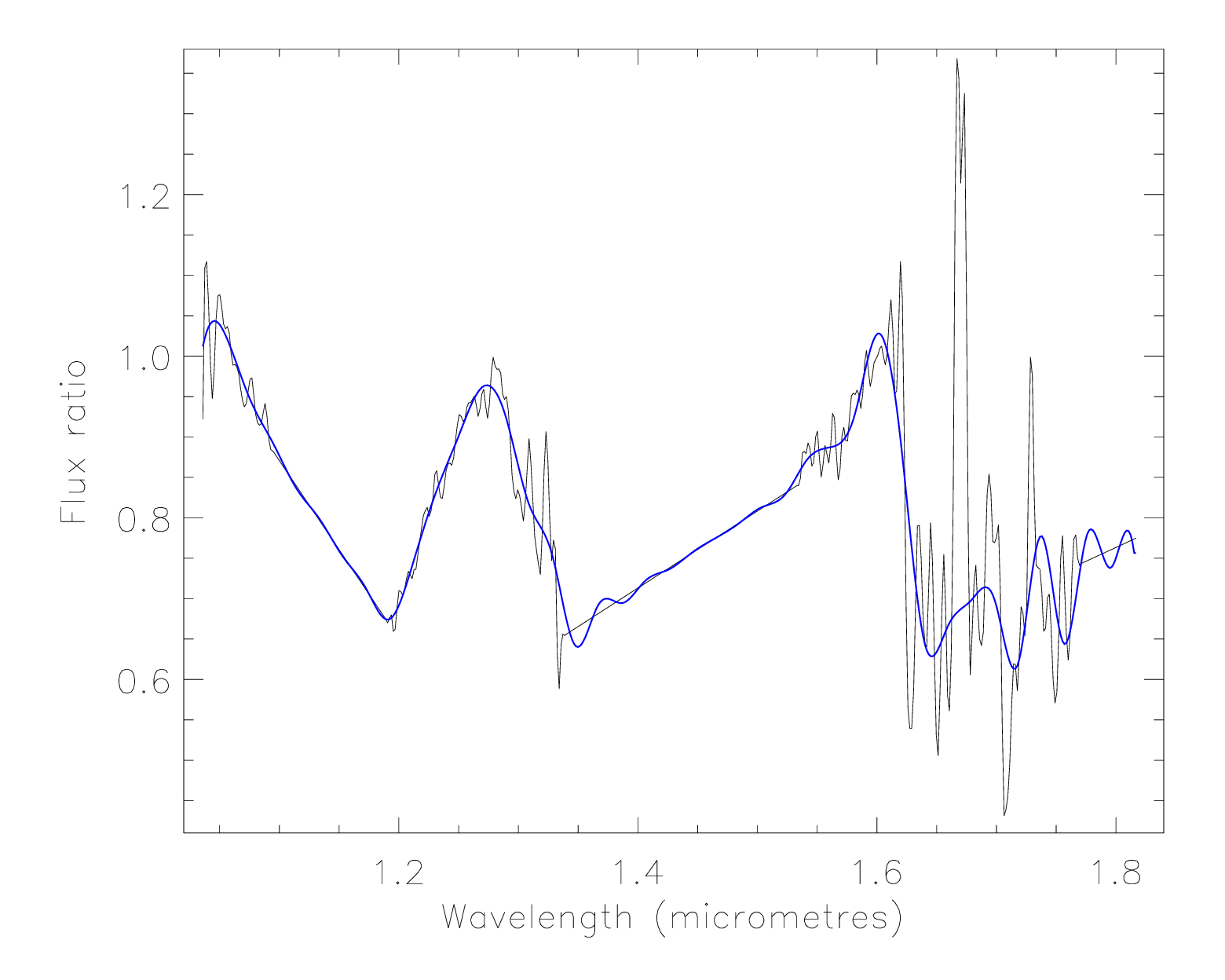}

\vspace{-4mm}
\includegraphics[scale=0.3,angle=0]{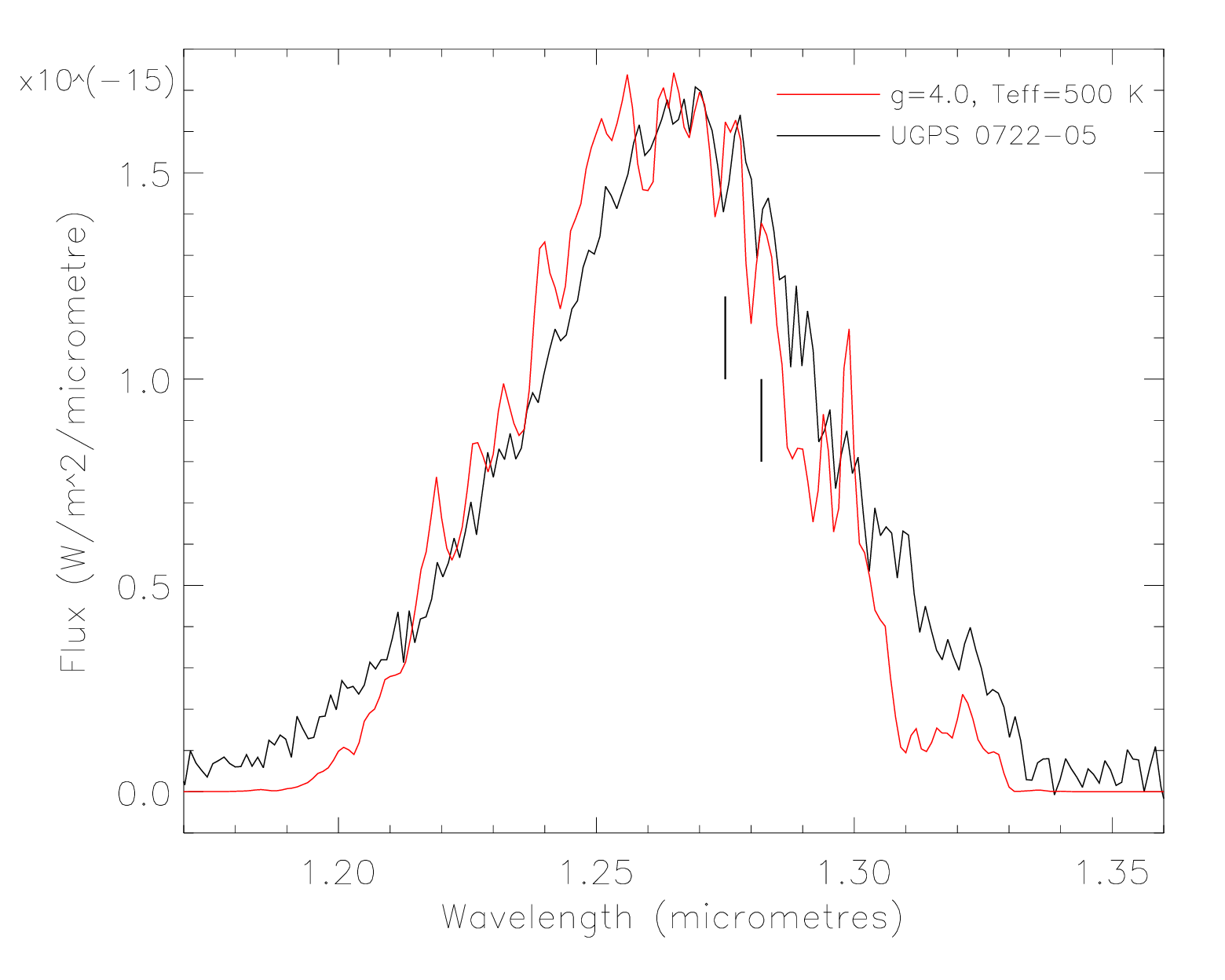}

\vspace{-4mm}
\caption{(upper panel) Near-infrared spectrum of UGPS 0722-05 (black line) and a T9 dwarf shown suitably
scaled and offset for comparison.
(middle panel) A lightly smoothed ratio spectrum (black line) showing the 1.05 to 1.8~$\mu$m region of the 
UGPS 0722-05 spectrum divided by the average of three T9 dwarf spectra and normalised to unity at 
1.279~$\mu$m. The blue curve shows a high order polynomial fit to the black line (the noise feature at 
1.67~$\mu$m was excluded from the fit). This plot shows clearly that the broad molecular absorption bands 
in UGPS 0722-05 are deeper on either side of the flux peaks near 1.27 and 1.59~$\mu$m, and on the long 
wavelength side 
of the 1.07~$\mu$m peak. Straight lines interpolate across noisy regions where there is 
little flux. (bottom panel) Expanded view of the $J$ band spectrum with a BT-SETTL model overplotted,
normalised to the same flux at 1.27~$\mu$m. Vertical lines mark the absorption features at 1.275~$\mu$m
and 1.282~$\mu$m.}
\end{figure}

\subsection{Spectroscopy}

The near-infrared spectrum of UGPS~0722-05 (Fig. 2, upper panel) is broadly similar to that of a T9 
dwarf. However, a ratio plot comparing UGPS~0722-05 with the average of three T9 dwarfs 
(Fig.~2, middle panel) shows that the broad molecular absorption troughs on either side of the flux peaks at 
1.28~$\mu$m and 
1.59~$\mu$m and on the long wavelength side of the 1.07~$\mu$m peak are between 10\% and 30\% deeper. 
The values of the spectral indices ``W$_J$'', ``CH$_4$-J'', ``NH$_3$-H'' and ``CH$_4$-H'', which are 
used for typing the coolest T dwarfs (see B08) are smaller than those of T9 dwarfs by 
amounts similar to the differences between T8 and T9 dwarfs, see Table 1.
We therefore assign a spectral type of T10. The expanded view of the
J band spectrum (Fig.~2, lower panel) shows a narrow absorption feature at 1.275~$\mu$m that
has an equivalent width of 3.6$\pm$0.1~\AA. A feature has been seen at a similar wavelength in 
Jupiter and (weakly) in a T8.5 dwarf and a T9 dwarf (see B08). We have examined a 
synthetic NH$_3$ spectrum generated from a new high temperature line list (see Yurchenko et al. 2009; 
Yurchenko et al.,in prep)
but there is no sign of a corresponding feature. The only candidate NH$_3$ absorption
feature in the spectrum is a weak detection in the $H$ band, at 1.514~$\mu$m 
(not shown). This is the wavelength of the strongest feature produced by a group of lines at 
1.4$<$$\lambda$$<$1.6~$\mu$m in the synthetic spectrum, when binned to the same resolution as 
the data. We caution that no conclusion can be drawn from this comparison until the line list is 
incorporated into a full model atmosphere.

In Fig. 2 (bottom panel) we overplot a BT-SETTL model atmosphere spectrum (see Allard et al. 2007), 
computed for effective temperature T$_{eff}$ = 500~K,
$g$=10$^4$~cm~s$^{-2}$ and [M/H]=0.0. Comparison with the data indicates that some marginally detected
narrow absorption features, e.g. at 1.282~$\mu$m, are probably real. Most of the narrow absorption 
features that appear in the model are due to H$_2$O (the NH$_3$ and CH$_4$ lists employed are highly 
incomplete in the $J$ band).
H$_2$O is a possible carrier of the 1.275~$\mu$m feature. Another is HF, which has some absorption 
lines at wavelengths close to this that are included in the models. This possibility was also suggested by 
Y.Pavlenko (private comm.). 
Despite the reasonable qualitative agreement between the model and the data in the $J$ band, the 
overall 1-2.5~$\mu$m SEDs predicted by all the BT-SETTL models at T$_{eff}$=400-600~K (not shown) 
are much bluer than we observe.

\begin{table}
    \caption{Astrometric solution for UGPS~0722-05}
    \begin{tabular}{lcccc}
Solution & No. of & $\mu$ & $\theta$ & $\pi$ \\ 
	 & epochs  	  & (mas/yr) & ($^{\circ}$) & (mas) \\ \hline
UKIRT only  & 7   & 972$\pm$8 & 291.2$\pm$0.2 & 237$\pm$41\\
UKIRT$+$2MASS & 8 & 967$\pm$8 & 291.1$\pm$0.2 & 246$\pm$33\\
  \hline
\end{tabular}
\end{table}

\vspace{-4mm}
\subsection{Parallax and proper motion}

\begin{figure}
\vspace{2mm}\includegraphics[scale=0.46,angle=0]{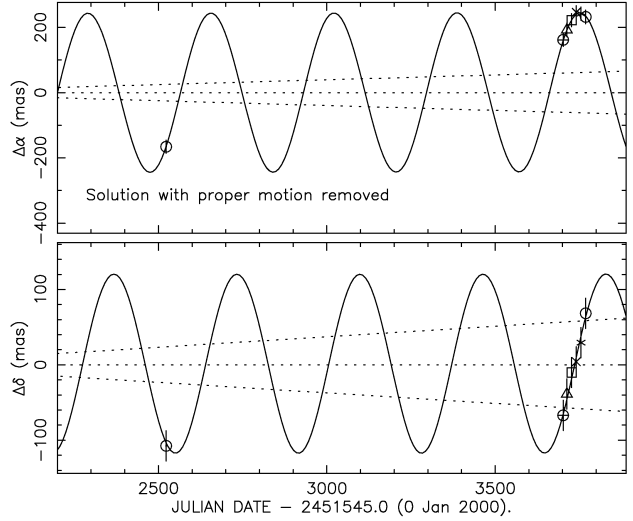}
\caption{The UKIRT$+$2MASS astrometric fit after proper motion (PM) subtraction.
Dotted lines show the PM uncertainties.
}
\end{figure}

An astrometric analysis was performed using the UKIRT and 2MASS data 
with a sample of 8 reference stars near 
UGPS~0722-05, using methods described in Tinney et al. (2003). The chosen stars
surround the target, range in brightness from 1.9 mag brighter
to 0.1 mag fainter than it (most are within 0.3 mag), and are found in both the 
2MASS and UKIRT images. 
The scatter of the residuals about the astrometric solution for reference stars 
within $\pm$0.2 mag of UGPS~0722-05 in UKIRT data is 6.2~mas. We adopt this as the astometric
precision for the T dwarf at each epoch.

The astrometric solution for UGPS~0722-05 using the UKIRT data alone clearly shows that it is 
nearby (see Table 3 -- the relative parallax is the weighted mean of those arising from the right 
ascension and declination solutions). 
However, the non-optimal 
sampling of its proper motion results in some concerns about degeneracy impacting on the parallax 
solution. Fortunately, the 2MASS datum helps us to refine the solution. Even though the 
observation has low signal-to-noise (the astrometric precision for the reference stars is $\sim$40
times lower than the UKIRT images) it adds in a much longer time baseline to constrain the proper motion. 
The UKIRT$+$2MASS solution (shown in Fig.3 for the 2006-2010 period) is consistent with the UKIRT-only 
solution (see Table 3), giving us confidence that the distance to this object has been well constrained.
At $d$=4.1$_{-0.5}^{+0.6}$~pc UGPS~0722-05 is the closest isolated brown dwarf known. Two even closer
stars with brown dwarf companions are known:
$\epsilon$~Indi (Scholz et al. 2003) and SCR~1845-6357 (Biller et al. 2006).
We note that neither the Gemini acquisition images (with 0.4\arcsec resolution) nor the {\it Spitzer} 
images 
reveal any sign of a companion.
A proper motion search for companions with the SuperCosmos archive also found nothing.

\section{Analysis and Discussion}

UGPS 0722-05 has absolute $J$ band magnitude, M$_J$=18.5$\pm$0.2, which is fainter than any 
other brown dwarf. Marocco et al. (in prep.) have measured parallaxes for a large sample of T dwarfs
and plotted M$_J$ vs. spectral type. Our assigned spectral type of T10 is consistent with the 
general trend that they report, within the scatter.

UGPS~0722-05 has $H$-[4.5]=4.71, which is redder than any other brown dwarf except the metal poor 
T7.5p dwarf SDSS~J1416+13B (Burningham et al. 2010b). The $H$-[4.5] colour is considered a good 
indicator of T$_{eff}$ (see e.g. L10a) so this supports our inference from the near-infrared 
spectrum that UGPS~0722-05 is cooler than the three known T9 dwarfs, for which 4.0$\le$$H$-[4.5]$\le$4.5. 
The {\it i-z}, {\it z-Y}, {\it Y-J}, {\it J-H} and {\it H-K} colours of the object are 
similar to those of the T9 dwarfs.
While the $H$-[4.5] colour is considered a good indicator of T$_{eff}$, it is also influenced by 
metallicity and gravity (L10a). Recent estimates of the distance to 
SDSS~1416+13B ($d\approx8$~pc, see Scholz 2010; Burgasser et al. 2010) indicate that it is more luminous 
than UGPS~0722-05 by a factor of $\sim$2. Assuming that these estimates are not greatly in error, 
they indicate that the redder colour of SDSS~1416+13B is due to a 
combination of low metallicity and high gravity, which would also explain the extremely blue 
{\it H-K} colour (see e.g. Leggett et al. 2009).
As Burgasser et al. (2010) pointed out, this 
implies that the T$_{eff}$ of SDSS~1416+13B is somewhat higher than the 500~K value that Burningham et 
al. (2010b) derived from the colours in the absence of a significant luminosity contraint. We therefore 
conclude that UGPS~0722-05 is the coolest brown dwarf known.

To calculate the total luminosity of the object we summed over the SED
as follows. The flux-calibrated near-infrared spectrum covers the range 1.035$<$$\lambda$$<$2.54~$\mu$m.
For the 0.94-1.035~$\mu$m we used the spectrum of the T9 dwarf ULAS~J003402.77-005206.7 (hereafter 
0034-00, Warren et al. 2007), 
scaled to the $Y$ magnitude of UGPS~0722-05. 
The negligible flux at $\lambda$$<$0.94~$\mu$m was not included.
For 3.92-4.00~$\mu$m we used the spectrum of the
T8 dwarf 2MASS~J04151954-0935066 (hereafter 0415-09, Saumon et al. 2007, hereafter S07), scaling 
to the 3.6~$\mu$m flux.
The fluxes in the 3.6~$\mu$m IRAC and T-ReCS $N$ (7.7-13.0~$\mu$m) passbands were calculated 
using the flux-calibrated spectra of 0415-09 (S07) and the T9 dwarf 
ULAS~J133553.45+113005.2 (hereafter 1335+11, Leggett et al. 2009) respectively. 
BT-SETTL models with T$_{eff}$=500-600~K and $log(g)$=4.0-5.0, [M/H]=0.0 were used for 
the 4.5~$\mu$m magnitude to flux conversion (lacking a suitable measured spectrum).
The same models were used to estimate the flux at $\lambda$$>$13.0~$\mu$m, at
2.54-3.18~$\mu$m and at 5.02-7.70~$\mu$m, scaling with the aid of fluxes in adjacent measured passbands.
The fluxes in these wavelength intervals have large uncertainties, owing to a strong dependence on
model parameters.

The total luminosity of UGPS~0722-05 is $L$=9.2$\pm$3.1$\times$10$^{-7}$~$L_{\odot}$, where the 
uncertainty arises from a 19\% uncertainty in the total flux and the 13\% uncertainty in the distance.
This compares
with $L$=1.1$\pm$0.1$\times$10$^{-6}$~$L_{\odot}$ for the T9 dwarfs 0034-00 and 1335+11 
(Marocco et al., in prep) and $L$=9.8$\pm$0.1$\times$10$^{-7}$~$L_{\odot}$ for the T8.5 dwarf
Wolf 940B (Leggett et al. 2010b), all of which have more fully measured SEDs
and much more precise parallaxes. 

Assuming an age in the range 0.2-10~Gyr, the evolutionary models of Saumon \& Marley (2008, 
hereafter SM08), allow a radius, $R$, between 
0.085~$R_{\odot}$ (at 10~Gyr) and 0.12~$R_{\odot}$ (at 0.2~Gyr). 
Using the definition of T$_{eff}$: $L$=4$\pi$$R^2$$\sigma$$T_{eff}$$^4$, we calculate 
T$_{eff}$=614$\pm$46~K for $R$=0.085~$R_{\odot}$ and T$_{eff}$=516$\pm$39~K for $R$=0.12~$R_{\odot}$.

Model atmospheres are not presently considered to be reliable at such low temperatures.
Nonetheless, some indication of physical properties
can be gained by consideration of evolutionary models and model atmospheres, aided by 
comparison with other late T dwarfs (see e.g. L10a). 
UGPS~0722-05 has {\it H-K} = -0.18$\pm$0.08 and M$_H$$\approx$18.8. In the top 
panel of fig. 9 of L10a it would lie at the bottom of the plot under Wolf 940B, indicating 
that the dwarf has similar gravity and metallicity to the T9 dwarfs 0034-00 and 
1335+11 but is significantly cooler, with T$_{eff}$$\approx$500~K.
Exactly the same conclusion can be drawn from placing it in the {\it H-K} vs. $H$-[4.5] diagram
in fig. 7 of Leggett et al. (2009).
The similarity otherwise to the three T8.5 to T9 dwarfs implies that UGPS~0722-05 has solar or slightly 
enhanced over solar metallicity. Its small tangential velocity ($\sim$19 km/s) suggests 
that it is not a very old object.

The luminosity calculation permits effective temperatures in the range 477-660~K (allowing for the
uncertainties in $L$ and $R$). However, similar
calculations for the T9 dwarfs 0034-00 and 1335+11 (which have well measured
luminosities), find T$_{eff}$=530-660~K for both objects (Marocco et al., in prep) and models indicate 
T$_{eff}<600$ K for 1335+11 (B08, Leggett et al. 2009; 2010a). The mature 
age-benchmark T8.5 dwarf Wolf940~B has T$_{eff}$$\approx$600~K (Leggett et al. 2010b), which is based on 
a well measured luminosity and a well constrained radius. Since UGPS~0722-05 is clearly cooler than 
these three objects, which have bluer $H$-[4.5] colours and earlier spectral types, we can be 
confident that its
effective temperature is in the lower half of the range permitted by the luminosity. We adopt
T$_{eff}$=480-560~K as the most likely range for this object. This is consistent with the 
500~K value indicated by the SM08 models. At 480-560~K, the SM08 models indicate a mass
in the range 5 to 15~$M_{Jup}$, $log(g)$ = 4.0 to 4.5 and age of 0.2 to 2.0~Gyr.

Whilst we have provisionally designated it as a T10 dwarf, we note that it is usual for the range of 
sub-types to run from 0-9. It is therefore quite possible that UGPS 0722-05 will come to be seen as the 
first example of a new spectral type. The fairly strong 1.275~$\mu$m feature might perhaps form the basis of 
such a classification. Gross changes in 1-2.5~$\mu$m spectra are not expected in cooler objects, given the 
similarity of T dwarf spectra to that of Jupiter (see B08) but the near-infrared 
component of the SED will decline.

\section{Acknowledgments}

The identification of the single good candidate late T dwarf found amongst several hundred 
million stars in the UKIDSS GPS is a tribute to the quality of UKIRT and the expertise of the staff at the 
Joint Astronomy Centre, the Cambridge Astronomical Survey Unit and the Wide Field Astronomy Unit at Edinburgh 
University. UKIRT is operated by the Joint Astronomy Centre on behalf of 
the Science and Technology Facilities Council of the UK. 
Gemini is is operated by the Association of Universities for Research in Astronomy, Inc., under 
a cooperative agreement with the NSF on behalf of the Gemini partnership, which consists of national
scientific organisations in the USA, the UK, Canada, Chile, Australia, Brazil and Argentina 
(see www.gemini.edu). The Subaru Telescope is operated by the National Astronomical Observatory of Japan.
This research has made use of the USNOFS Image and Catalogue Archive operated by the United States Naval 
Observatory, Flagstaff Station.\\

{\large \bf References}\\
Artigau E., et al., 2010, arXiv:1006.3577\\
Allard F., et al., 2007, A\&A, 471, L21\\
Biller B.A., et al., 2006, ApJ, 641, L141\\
Burgasser A.J., Kirkpatrick J., et al., 2002, ApJ, 564, 421\\
Burgasser A.J., et al., 2010, AJ, 139, 2448\\
Burningham B., et al., 2008, MNRAS, 391, 320 (B08)\\
Burningham B., et al., 2010a, MNRAS (in press)\\
Burningham B., et al., 2010b, MNRAS, 404, 1952\\
Costa E., et al., 2006, AJ, 132, 1234\\
Delorme P., Delfosse X., et al., 2008,  A\&A, 482, 961\\
Lawrence A., et al., 2007, MNRAS, 379, 1599\\
Leggett S.K., Geballe T., et al., 2000, ApJ, 536, L35\\
Leggett S.K., et al., 2009, ApJ, 695, 1517\\
Leggett S.K., et al., 2010a, ApJ, 710, 1627 (L10a)\\
Leggett S.K., et al., 2010b, ApJ (submitted)\\
Lodieu N., Dobbie P.D., et al., 2007, MNRAS, 380, 712\\
Lucas P.W., et al., 2008, MNRAS, 391, 136 (L08)\\
Luhman K.L., 2004, ApJ, 617, 1216\\
Saumon D., et al., 2007, ApJ, 656, 1136\\
Saumon D., Marley M.S., 2008, ApJ, 689, 1327\\
Scholz, R.-D., et al., 2003, A\&A 398, L29\\
Scholz R-D., 2010, A\&A, 510, L8\\
Skrutskie M.F., Cutri R.M., et al., 2006, AJ, 131, 1163\\
Smart R., et al., 2010, A\&A, 511, A30\\
Tinney C.G., 1998, MNRAS, 296, L42\\
Tinney C.G., et al., 2003, AJ, 126, 975\\
Warren S.J., Mortlock D.J., et al., 2007, MNRAS, 381, 1400\\
Weights D.J., Lucas P.W., et al., 2009, MNRAS, 399, 2288\\
Yurchenko S.N., et al., 2009, Phys. Chem. A., 113, 11845

\end{document}